**Echo Chambers Exist! (But They're Full of Opposing Views)**

Jonathan Bright, Nahema Marchal, Bharath Ganesh, Stevan Rudinac

**Correspondence:** jonathan.bright@oii.ox.ac.uk

**Abstract:**
The theory of echo chambers, which suggests that online political discussions take place in conditions of ideological homogeneity, has recently gained popularity as an explanation for patterns of political polarization and radicalization observed in many democratic countries. However, while micro-level experimental work has shown evidence that individuals may gravitate towards information that supports their beliefs, recent macro-level studies have cast doubt on whether this tendency generates echo chambers in practice, instead suggesting that cross-cutting exposures are a common feature of digital life. In this article, we offer an explanation for these diverging results. Building on cognitive dissonance theory, and making use of observational trace data taken from an online radical right website, we explore how individuals in an ideological "echo chamber" engage with opposing viewpoints. We show that this type of exposure, far from being detrimental to radical online discussions, is actually a core feature of such spaces that encourages people to stay engaged. The most common "echoes" in this echo chamber are in fact the sound of opposing viewpoints being undermined and marginalized. Hence echo chambers exist not only in *spite of* but *thanks to* the unifying presence of oppositional viewpoints. We conclude with reflections on policy implications of our study for those seeking to promote a more moderate political internet.

**Keywords:** echo chambers, cognitive dissonance, online politics, extremism, selective exposure

**Acknowledgments:** Earlier versions of this paper were presented at the Oxford Internet Institute brown bag seminar and the ICS Lisbon Sparc Seminar series. The authors are grateful for insightful feedback and comments from audience members.

**Funding:** The work was financed by the VOX-Pol Network, which is funded by the EU 7th Framework Programme (grant number 312827).

**Introduction**

The theory of echo chambers suggests that online political discussions tend to take place in conditions of ideological homogeneity (Sunstein, 2001). Variants of this theory have existed since the early days of computer-mediated communication research (Negroponte, 1995; Papacharissi, 2011; Van Alstyne & Brynjolfsson, 2005). At the micro-level, strong support has been found for behavioral characteristics such as selective exposure and homophilous sorting, which suggest that individuals should tend to arrange themselves in ideologically consonant groups (see e.g. Garrett, 2009; Iyengar and Hahn, 2009; Knobloch-Westerwick and Meng, 2009). However, when people's information environments are considered holistically, the picture is much more mixed. Survey evidence has repeatedly shown that individuals — especially the most politically engaged — are frequently exposed to cross-cutting information and opinions online (Dubois & Blank, 2018; Flaxman, Goel, & Rao, 2016; Fletcher & Nielsen, 2018) with vitriolic exchanges between political rivals a common feature of digital life. Macro-level studies that examine connections at a network level based on digital trace data also find strong patterns of exchange between those of different ideological persuasions (Bright, 2018; Gentzkow & Shapiro, 2011). Thus, while online information environments rarely follow a random distribution of viewpoints, evidence seems to suggest that most Internet users, even extreme ones, are anything but sheltered from challenging views. One explanation for this divergence between micro and macro level findings is that we do not know enough about how individuals in (largely) ideologically homogeneous spaces react to exposure to opposing viewpoints. The behavioral consequences of this exposure have been under researched, especially outside of a lab setting. This gap in the literature is especially important because much of the academic and policy literature has started to promote the idea of "breaking free" of echo chambers as a means of decreasing polarization and increasing political tolerance (Bessi et al., 2016; Sunstein, 2017; Zollo et al., 2017). Our article addresses this gap in the literature. The research question is simple: How do individuals within an echo chamber react when exposed to opposing viewpoints? To address this question, we make use of observational trace data drawn from a well-used white nationalist online forum, Stormfront. This forum presents an excellent case study for our question as it possesses many characteristics of an ideological "echo chamber", yet also frequently hosts opposing views. The article is structured in the following way. In the next section, we outline the theory of echo chambers, grounding it in literature on cognitive dissonance. We then derive a number of hypotheses from cognitive dissonance theory, and the work which has followed it, regarding how individuals might respond when exposed to opposing viewpoints. Then, we describe more fully our case study and the nature of our data and measures. Finally, we present our results. We show evidence that, far from being an anathema, opposing views appear to be a core part of an echo chamber. Members of the echo chamber actively engage with them, working to weaken and reframe these views such that they bolster the overall ideology espoused by the forum. These results are discussed in terms of their consequences for theory on radicalization and polarization, and efforts to inject more cross-cutting exposure into online life.

**Echo Chambers and Exposure to Dissonant Material**

Following previous work in this area, we define an echo chamber as an online social discussion space such as a forum, blog or social networking site which is characterized both by relative *ideological homogeneity* in terms of information being circulated and opinions being expressed (Sunstein, 2001; Van Alstyne & Brynjolfsson, 2005) and by relative *extremism* of those opinions, in the sense that they diverge from broader societal norms in the area. Both of these elements are worth reviewing briefly here.

Ideological homogeneity means, simply, that the majority of people in a discussion space have largely similar viewpoints. Most work on echo chambers has explained their ideological homogeneity on the basis of the theory of cognitive dissonance (for examples see Garrett, 2009, pp. 267-268; Colleoni, Rozza and Arvidsson, 2014, pp.318-319; Hayat and Samuel-Azran, 2017, p. 294; Dubois and Blank, 2018, p. 731). As Festinger, who first developed the theory, puts it: "the individual strives towards consistency" in their thoughts and beliefs (1957, p. 1). When individuals encounter inconsistencies, which Festinger labels "dissonances", between their beliefs and other attitudes or opinions they become "psychologically uncomfortable" (1957, p. 3). This discomfort motivates people to act to eliminate or reduce the dissonance they are experiencing. Festinger argued that dissonance reduction was "a motivating factor in its own right" (*ibid*), something essential which stimulates action, similar to hunger and thirst. The considerable follow-up literature in this area has debated this claim (for discussions see Brehm, 2007; Harmon-Jones and Harmon-Jones, 2007; Fiske, 2010, pp. 253-254; Gawronski, 2012); and some have proposed alternative sources of motivation such as the desire to minimize mental load (Fischer, 2011), perceptions of untrustworthiness in attitude-discrepant information sources (Metzger, Hartsell, & Flanagin, 2015) or the motivation to protect feelings of self-worth (Bem, 1967; Harmon-Jones, Haslam and Bastian, 2017, p. 2). However, regardless of the precise source of the motivation, the basic tenets of dissonance theory have been repeatedly validated in a wide variety of studies (Elster, 2007, p. 19). Specifically, there is good evidence that people experience psychological discomfort when exposed to dissonant information (see e.g. Zanna and Cooper, 1974; Elliot and Devine, 1994; Harmon-Jones *et al.*, 1996). and engage in practices aimed at reducing such dissonance as a result (Gibbons, Eggleston, & Benthin, 1997; Kiesler & Pallak, 1976; Losch & Cacioppo, 1990). One of these practices is to seek out ideologically consonant spaces and interactions, which would explain the emergence of ideologically homogeneous echo chambers.

The type of ideological homogeneity observed in echo chambers can, in turn, breed extremism and even violence (Stroud, 2010; Wojcieszak, 2010; Ganesh and Bright, 2020a). Research has shown that interpersonal discussion between like-minded others increases certainty that beliefs are correct (Frey, 1986; Sunstein, 2001), thus reinforcing attitude extremity (Bennett & Iyengar, 2008; Iyengar & Hahn, 2009; Stroud, 2010). With this certainty comes greater intolerance towards groups who hold differing opinions (R. Kelly Garrett, Weeks, & Neo, 2016). Individuals who actively participate in ideologically extreme online groups tend to believe that their opinions are superior and more prevalent than rival ones (Wojcieszak, 2010). Furthermore, homogenous groups are particularly prone to normative and informational pressures (Deutsch & Gerard, 1955): when group members adopts a particular viewpoint or behavior, such as aggressivity, as the norm, others follow suit (Hogg, Turner, & Smith, 1984). Likewise, exposure to only in-group viewpoints can increase the extent to which individual's self-identities are founded on these groups, resulting in a tendency to favor one's own social group and denigrate others (Mackie, Devos, & Smith, 2000; Smith, Seger, & Mackie, 2007). At the most extreme ends, this "segregation" of beliefs has been shown to foster political hatred and violence (Melton & Motyl, 2018).

There is, therefore, real concern with the radicalizing potential of online echo chambers. In recent years, this has led many to suggest that echo chambers were "destroying" democracy, and led to growing calls to break out of one's "filter bubble" (Pariser, 2011). Rooted in a deliberative tradition that sees exposure to a diversity of opinion as normatively desirable, government programs aimed at fighting radical online extremism have also tended to favor "counter-narratives" and other counter-messaging techniques (Rosand and Winterbotham, 2019; Ganesh & Bright, 2020b). Though laudable, these efforts rest on the double assumption that online echo chambers are hermetic, and that injecting these ideological silos with opposing viewpoints will generate positive outcomes in the long run. Currently there is not enough empirical evidence about how individuals in an echo chamber react to dissonant views to support these claims. Our core interest in this article lies in understanding the impact of this type of engagement, particularly for people who have already made some effort to enclose themselves in an ideologically homogenous discussion group. Our research question is simple:

*RQ: How do individuals within an echo chamber react when exposed to opposing viewpoints?*

The theoretical basis of our article is work around the concept of cognitive dissonance. As we have highlighted above, this is the foundation on which most studies of echo chambers rest. Dissonance theory makes a number of propositions about how individuals ought to react when exposed to dissonance. Here, we divide these into two groups, the first relating to the extent and nature of engagement with opposing viewpoints, and the second relating to longer term behavioral consequences for those who engaged.

We will begin with our question on engagement. Most online discussion spaces, including the one in our study, have a post and comment structure (Gonzalez-Bailon, Kaltenbrunner, & Banchs, 2010). That is to say, individuals using a forum create new posts containing original content, to which other users of the forum respond with comments, forming what are often called discussion threads. It may be that, in an ideological echo chamber, dissonant viewpoints receive less comments, and hence generate shorter discussion threads, than consonant ones. Clearly, people using a forum have free choice in terms of which discussion threads they participate in. And, as we have highlighted above, the theory of cognitive dissonance suggests that members of an echo chamber might avoid discussion threads containing dissonant viewpoints: Festinger argues that the most intuitive strategy to reduce cognitive dissonance is to avoid situations that are likely to produce dissonance in the first place (see Festinger 1957, p. 3, see also pp. 129-131). There is a considerable body of evidence supporting this: people not only tend to selectively avoid information they disagree with online (see e.g. Smith, Fabrigar and Norris, 2008; Iyengar and Hahn, 2009; Knobloch-Westerwick and Meng, 2009), but sometimes take active measures to avoid dissonant views altogether, such as unfriending or unfollowing others (John & Dvir-Gvirsman, 2015; Yang, Barnidge, & Rojas, 2017). Support for this idea is also found in the closely related theories of "selective exposure" and "selective avoidance" (for reviews see O'Keefe, 2002; Stroud, 2014; Knobloch-Westerwick, 2015).

However, we can also imagine that dissonant material may generate more responses than consonant material. That is because, in addition to individual level motivations, dissonance theory is also a theory of group behavior (Matz & Wood, 2005). As Festinger puts it, social groups are "at once a major source of cognitive dissonance…and a major vehicle for eliminating and reducing [it]" (1957, p. 177). When members of the same group collectively encounter dissonant beliefs in a social environment, they may seek to contradict and undermine them; a process referred to as "trivialization" by Simon, Greenberg, & Brehm (1995) or "denial" by Abelson (1959, pp. 344-345). This trivialization can involve direct counter-argument, undermining the logical basis on which the dissonant cognition rests, such that the total line of reasoning is recast as invalid (Festinger 1957, p. 135). Another way of achieving this is to undermine the status of the speaker. As Festinger put it, if the source of the dissonant cognition can be cast as "stupid, ignorant, unfriendly and bigoted" then the extent of dissonance is reduced (1957, p. 183). It is therefore reasonable to assume that intergroup processes will lead forum members to want to maintain and even reinforce their beliefs in the face of such dissonance, which could mean that dissonant material is more likely to attract a response. Indeed, when members of a group all experience dissonance at the same time, this may trigger cascading effects whereby any responses to this dissonance are enhanced by social support provided by others also trying to minimize their own dissonance (see Festinger, 1957, pp. 196-202). On the basis of these diverging expectations, we propose the following hypothesis:

*H1: Dissonant posts will generate less engagement than consonant posts on the forum*

In addition to the amount of responses generated, we might also expect different types of people to engage with dissonant material. Work in this area shows that the level of dissonance people experience when having their beliefs challenged is moderated by the strength of their original viewpoint (Lavine, Lodge, & Freitas, 2005) and how closely the issue ties in to their self-concept (see e.g. Aronson and Mettee, 1968). In other words, the more strongly someone identifies with a particular group or ideology, the more likely they are to react to challenges. There is good evidence that people with different political orientations or dispositions, for example, vary in their willingness to associate with likeminded others, with politically extreme individuals more likely to be homophilous in their associations (Boutyline & Willer, 2017; Jost, van der Linden, Panagopoulos, & Hardin, 2018). According to social identity theory, people derive part of their self-concept from membership of distinct real or virtual social groups formed around shared interests, identities or values (Michael A. Hogg, Abrams, Otten, & Hinkle, 2004). In this view, long-term membership and active participation in an online community such as a political forum is not only testament to someone's ideological dispositions, but also to the strength of their political identity (Carr, 2017). In our particular case, we might therefore expect users who are more long time and established members of the forum to have a heightened sense of in-group identity and to react more strongly than newcomers when encountering dissonant material. We thus elaborate our next hypothesis:

*H2: More senior members of the forum will be less likely to engage with oppositional material*

We will now move on to discussing potential behavioral consequences of engaging with dissonant material. Despite the possibility of selective avoidance mentioned above, Festinger felt that some engagement with dissonant viewpoints was inevitable, because "a person does not have complete and perfect control over the information that reaches him and over events that can happen" (1957, p. 4). In the context of a forum, one possible behavioral response to exposure to dissonant material would be to cease participation, perhaps because the effort of simply ignoring disagreeable contributions already creates an uncomfortable amount of cognitive dissonance. Indeed, Festinger felt that, as the amount of dissonance increases, this type of avoidance behavior grows stronger (see 1957, p. 130). In this perspective, over the long term, the frequent appearance of oppositional voices should cause echo chambers to break down entirely, as people gradually leave the conversation. Supporting evidence for this idea has been presented by Matz & Wood (2005), who demonstrated how people changed social groups when faced with intra-group dissonance (2005, pp. 34-35). It is also supported by literature on political disengagement: for example, Torcal and Maldonado (2014) have shown how exposure to political disagreement and argument can lead to people dropping out of politics. Survey research confirms his: a great majority of social media users find it "stressful" to talk politics online with people they disagree with, and many ignore or simply "move on" from dissonant content when they encounter it (Duggan & Smith, 2016). We thus develop our third hypothesis:

*H3: After engaging with dissonant views, users will be more likely to abandon a forum*

A further possible behavioral response is that users may seek to change their internal information balance by counteracting the opposing view with other confirmatory views, perhaps greater in number. As Festinger argued, dissonance can be reduced by adding new evidence which confirms the original viewpoint and hence dilutes the impact of the original oppositional voice (Festinger, 1957, p. 22). This line of thinking has been supported by further work, which has shown how people become more likely to seek attitude consonant cognitions when presented with dissonant information (Cohen, 1960; Eagly & Chaiken, 1998; Ehrlich, Guttman, Schönbach, & Mills, 1957; Frey, 1986).

Group processes may be especially important for adding consonant cognitions, because these supportive pieces of information are boosted by the social support provided by group members (Festinger, 1957, p. 179). Indeed, research has shown that membership of radical groups allows individuals to reduce the dissonance involved in holding radical views (McKimmie et al., 2003). This is because even though these views might be generally in the minority, in the local context of an echo chamber they form the majority (Cooper & Stone, 2000). And, "given high levels of certainty or confidence, exposure to discrepant information may not be dissonance arousing" (Stroud, 2010, p. 560). Hence, at the individual level, we might expect those coming into contact with dissonant content to seek to consume confirmatory content shortly afterwards, perhaps by engaging with consonant material or seeking out consonant areas of the site. This leads us to our final hypothesis:

*H4: Individuals will seek out attitude consonant material after engaging with dissonant material*

**Data and Measures**

**Data**

We draw our data from the website Stormfront, which is a major online forum white nationalists with a long history of hosting extreme discussion (Caren, Jowers, & Gaby, 2012; De Koster & Houtman, 2008; Hara & Estrada, 2003) and promoting radical viewpoints which are far from the societal mainstream (see Meddaugh and Kay, 2009, for a fuller analysis of the viewpoints circulating on the site). Stormfront has a classic forum structure: it is powered by vBulletin, a widely used piece of forum hosting technology. It is divided into a number of thematic sub-forums, where users can discuss issues of their choice. The discussion itself takes a post-comment format: people using the site can propose new discussion topics ("posts"), and then other people comment on these posts. The post and its comments form a discussion thread.

We chose this forum for study for two reasons. First, the site fits the concept of an echo chamber which we have defined above. It is largely ideologically homogeneous, and it contains within it extreme viewpoints, which are marginalized in mainstream society but which are in a clear majority on the forum (indeed, one of the major reasons that users go there is to hear affirmation of viewpoints which they cannot access in their daily lives – see Bowman-Grieve, 2009). However, the site is also not a *complete* echo chamber. In fact, there is a special sub-forum within the site dedicated to hosting "opposing views", where people can promote ideas which are dissonant with the idea of white nationalism in some respect. The existence of this part of the forum allows us to address our question of how members of an echo chamber react when exposed to opposing viewpoints.

The data used for the paper was harvested automatically from the forum using a web spidering technique. The spider collected both the text of posts contained in the forum, and any comments or which were made as part of the post's discussion thread. It also collected the username of accounts making posts and comments. The data used in the study was harvested from seven sub-forums: the opposing views sub-forum as described above, and six other sub-forums which do not permit oppositional discussions. These other six sub-forums were selected to provide a range of different types of discussion topic (from leisure and dating to politics and activism), and to be broadly similar in size to the opposing views sub-forum. In total we collected just over 280,000 contributions to the forum. Data was only collected from open access parts of the forum, where no account was required to access the material, and has only been presented in an aggregate pattern to obscure the identity of the participants. The dataset which is released to accompany this paper is similarly de-identified. Further notes on sub-forums and data are available in appendix A1.1. The dataset collected covers the period 2011 – 2013. Post-hoc verification of the data collected by the web spider suggested it was approximately 93.2% complete (see appendix A1.2 for details of the verification process).

**Measures**

In this section we outline our measures (descriptive statistics for all variables are available in appendix A1.5). Our study concerns how members of an echo chamber react to exposure to dissonant viewpoints. A first critical measurement task was, therefore, to characterize the type of engagements users had during their time on the forum. These engagements occur when users make a post or comment within a particular sub-forum. We define three such engagement types. First, users could engage in discussions by commenting in one of the ideologically consonant sub-forums (i.e. any sub-forum apart from the opposing views sub-forum). These sub-forums are exclusively for discussion amongst white nationalists, and dissonant material is actively deleted by forum moderators. Second, users could engage with "consonant arguments" inside the opposing views sub-forum. These are discussion threads in the opposing views sub-forum where the initial post was in support of white nationalism. Finally, users could engage in "dissonant arguments" in the opposing views sub-forum: discussion threads where the initial post was against white nationalism.

We distinguished between consonant and dissonant arguments within the opposing views sub-forum by manually coding each post within this forum — that is, each original post to the forum that started a new discussion thread — into one of three categories: consonant, dissonant, and miscellaneous. Consonant posts were ones that agreed with and supported the major tenets of white nationalism (our interpretation of white nationalism is described in appendix A1.3). Dissonant posts were ones that contradicted this belief system, and everything else was in the miscellaneous category. Krippendorff's alpha was calculated as 0.814 for the coding exercise, with a percentage agreement of 88%. The posts in general were quite long and required considerable time to read and code appropriately. Hence, we chose to code only contributions from the year 2012 in our data. We decided to use a continuous time period rather than a random sample of data because this facilitated the gap time and Markov chain analyses we chose to employ (these analyses are described further below). In total 1,468 posts were coded, which together attracted 34,368 comments. Further details on the coding process are provided in appendix A1.4.

A second measurement task was to identify who counts as a "member" of the Stormfront echo chamber, especially when the forum contains contributions from people with views opposed to white nationalism. We approached this task in three ways. First, we discounted "guest" users, and considered only those people who had registered an account on the site. Second, we discounted people who had never made a contribution outside of the opposing views sub-forum: having contributed in one of the other ideologically consonant sub-forums is, we would argue, a good indicator of being a member of the echo chamber. Finally, of course, we discounted anyone who made a dissonant post within the opposing views sub-forum itself.

We also collected a number of other pieces of information. We measured the seniority of each user in our dataset, based on the amount of previous contributions they had made to the forum before their current engagement. We use our data from 2011 in this calculation as well. We measured the length of the post, defined as the number of characters, to account for any potential confounding effects this might have on our main relationship of interest (for example, dissonant viewpoints might be longer ones, and longer ones might inherently be more likely to generate responses). And, for each engagement, we observed the location of a user's next engagement, using data from 2013 if necessary. Descriptive statistics on all measures collected can be found in appendix A1.5.

**Limitations**

It is worth acknowledging the limitations to our study which are inherent in the data we have collected. The major limitation is our focus on just one website, and just one time period. This narrow focus was necessitated by the fact that the qualitative coding task was nuanced and time consuming, and also that a time series of observations was necessary to address our hypotheses (as we will describe further below). However, further work could usefully address whether our findings generalize to other contexts. A second issue is that we focus only on *active* engagements in the Stormfront forum, as these are the only things we have data on. We cannot know how many people actually *read* the contributions we are interested in, nor what the impact of reading (as opposed to actively engaging through commenting) was. Again, this would be an interesting topic for further work.

Third, there is the issue of moderation. Although in theory anyone is free to create an account and to create content, we know that in practice the forum receives a lot of moderation — something which was referred to frequently in the contributions that we read. Therefore, the oppositional viewpoints which make it onto the forum do not represent the full range of all possible opposition, but rather the ones that the moderators have selected for inclusion, or at least not removed after the fact. We know that the moderators are somewhat strategic in this behavior, and often mention that they are "allowing" opposing viewpoints in so that key arguments against them can be rehearsed. We will return to this point in the conclusion. Regardless of the moderation, however, we do have sufficient presence of oppositional viewpoints to address our hypotheses.

Finally, there is the issue of how seriously people hold their views. As has been repeatedly been documented, online spaces such as forums are places where people can play and experiment with their identities (Hogan, 2010; Marwick, 2005; Marwick, 2013). It may be that many of the 'members' of the forum don't actually feel passionately about white nationalism; and conversely it may be that many of the dissonant voices are simply trolling or acting in an ironic fashion (Nagle, 2017). We cannot account for this in our data. However, we would point out that a considerable body of existing research has argued that Stormfront contains many people who genuinely believe in what they are saying, and has argued for the importance of the site (see for example Bowman-Grieve, 2009).

**Analysis**

We begin our analysis by assessing evidence for our first and second hypotheses, that opposing viewpoints will prompt less engagement from members of the forum (H1), and they will be less likely to generate engagement from people who are long-term members of the forum (H2). We address this in a series of multilevel models reported in Table 1. Each observation in these models is a post made to the forum in the year 2012. We include the anonymized username of the initial poster, the day of the week and the month of the year as random effects. Estimates of statistical significance were computed using the Kenward-Roger approximation for degrees of freedom (Halekoh & Højsgaard, 2014). Marginal and conditional $R^2$ were calculated using the method proposed by Nakagawa and Schielzeth (2013). Diagnostics show no reason to doubt the results reported in the models (information on diagnostics can be found in appendix A2.1).

Models 1.0 to 1.2 address our avoidance hypothesis (H1), looking at whether dissonant material receives fewer responses than consonant material. Model 1.0 simply compares all material found in the opposing views forum to all material found outside it. While the opposing views posts do seem to generate more responses, the difference is not significant. M1.1 breaks down the opposing views content into three categories, as defined above: consonant arguments, dissonant arguments and miscellaneous. Dissonant material again generates more responses than content outside the opposing views forum, but the result is only borderline significant. Model 1.2, finally, focusses only on material within the opposing views sub-forum. Here there is a significant effect: dissonant arguments receive more than twice the amount of responses

**Table 1: Analytical Models of Post Response Levels and Contributor Seniority**

|  | Amount of responses (log) | | | Previous contribution levels of responders (log) | | |
| --- | --- | --- | --- | --- | --- | --- |
|  | M1.0 | M1.1 | M1.2 | M2.0 | M2.1 | M2.2 |
| OV Forum Post | 1.49 | | | 3.68 *** | | |
| OV Consonant | | 1.21 | | | 3.71 *** | |
| OV Dissonant | | 2.74 + | 2.29 *** | | 3.70 *** | 1.05 |
| OV Misc | | 0.94 | 0.78 ** | | 3.52 *** | 0.90 |
| Content length (log10) | 1.12*** | 1.09*** | 1.04 | 1.09 *** | 1.09*** | 0.99 |
| Marginal $R^2$ | 0.03 | 0.06 | 0.09 | 0.22 | 0.22 | 0.00 |
| Conditional $R^2$ | 0.22 | 0.20 | 0.23 | 0.24 | 0.24 | 0.19 |
| Observations | 4,264 | 4,264 | 1,468 | 3,353 | 3,353 | 1,198 |

*Note: * p < 0.05; ** p < 0.01; *** p < 0.001*

that consonant arguments do. Hence overall there is some evidence that opinion dissonant material generates more reaction than consonant material does amongst members of an echo chamber. Importantly, there is no evidence that it generates less reaction. Hence H1 is not supported.

Models 2.0 to 2.2 address our next hypothesis (H2), that people with more deeply held beliefs should be less likely to contribute to dissonant discussions. In these models, each observation is a post and the dependent variable is the average seniority of the people replying to that post (measured as the number of contributions they have previously made to the forum). There is clear evidence against the hypothesis. Model 2.0 shows that people responding within the opposing views forum are more than three times as senior as those responding outside.

Model 2.1 shows that this is true for all types of opposing views forum posts. Model 2.2 shows that, however, the differences *between* different types of opposing views post are not significant.

**Table 2: PWP Gap Time Models**

|  | Probability of posting again | |
| --- | --- | --- |
|  | M3.0 | M3.1 |
| OV Forum Post | 1.18 *** |  |
| OV Consonant |  | 1.20 *** |
| OV Dissonant |  | 1.14 *** |
| OV Misc |  | 1.28 *** |
| Number of Replies (log10) | 1.16 *** | 1.16 *** |
| $R^2$ | 0.012 | 0.013 |
| Users | 3,324 | 3,324 |
| Events | 57,625 | 57,625 |

*Note: * p < 0.05; ** p < 0.01; *** p < 0.001*

We will now move on to our third hypothesis, that of abandonment (H3). Forum users may not actively avoid oppositional viewpoints, but does engaging with them make them less likely to use the forum in the future? In Table 2 we address the question making use of a recurrent events model (specifically, we use a PWP Gap Time model – see Amorim & Cai, 2015; Prentice, Williams, & Peterson, 1981). Such an analysis is suited to our case because we do not observe whether (or when) people definitively decide to stop contributing to a forum, just the amount of time between each post. The technique also handles censoring in our data.

Model 3.0 looks at the impact of engagements within the opposing views forum compared to engagements outside of it. There is good evidence against H3: people using the opposing views forum are about 17% more likely to return than those engaging outside of it. Model 3.1 breaks this down by the type of opposing engagement. It shows that while all types of engagement have a positive effect, engaging with consonant arguments has more of an effect than engaging with dissonant arguments. Diagnostics show that the proportional hazards assumption of both models is violated, indicating the result changes over time, with the strongest part of the effect coming in the first 80 minutes after the initial post (see appendix A2.2 for more details on these diagnostics). Hence encounters with dissonant viewpoints stimulate a brief but immediate likelihood of returning to the forum, contradicting H3.

We will now move on to our final hypotheses (H4), tackling re-confirmation. Given opposing viewpoints do not stimulate either avoidance or abandonment responses (indeed, quite the opposite), do they prompt people to seek consonant material after exposure? We tackle this question by making use of a discrete time Markov chain model. In the model, we treat engagement with the different categories of content we identify (consonant arguments, dissonant arguments, and discussions outside of the opposing views forum) as different states between which people transition, with the transitions indicated by their engagements with the forum. These records of interactions are used to infer the transition probabilities in the model, following the method proposed by Yalamanchi and Spedicato (2015). If our reconfirmation hypothesis (H4) is correct, there should be a higher probability of transitioning outside of the opposing views forum when encountering a dissonant argument than there is when encountering a consonant one.

**Table 3: Markov Chain Transition Probabilities**

|  | Probability of transition outside of opposing views forum | |
| --- | --- | --- |
| Having seen a… | Estimate | 95% Confidence Interval |
| Consonant viewpoint | 20.45% | 19.54% - 21.35% |
| Dissonant viewpoint | 22.99% | 22.08% - 23.89% |
| Number of users | 3,324 | |
| Number of state transitions | 57,625 | |

The inferred transition probabilities and associated confidence intervals are shown in Table 3 (full transition probabilities for the model are available in appendix A2.3). The results support H4: there is evidence that people engaging with dissonant viewpoints are around 2.5 percentage points more likely to transition outside of the OV forum for their next engagement than people who engage with consonant material (another way of expressing this result, which is similar to how it would be expressed through a logistic regression, is that the chances of transition outside of the opposing views forum increase by around 12% for those encountering dissonant viewpoints).

As a final analytical step, to complement the quantitative data above, we will seek to describe in more detail the type of discussions which emerged around these dissonant viewpoints on Stormfront. While performing the coding process, we engaged in a qualitative analysis of the types of responses to dissonant posts, following a similar process to that of Yardi and boyd (2010). This process was inductive in that it was conducted without prior expectations of the types of responses we would find: all themes and counter-argument strategies identified were grounded in the data itself. After coding was completed, this was worked into a typology of dissonant engagements. Combining quantitative and qualitative analysis in such a way allows us to better understand the perspective of the forum members themselves.

First, we identify *repeated rebuttal* as a common counter-argument type. Dissonant viewpoints that appear on the site, especially those that strongly criticize white nationalism on the grounds of either racism, hypocrisy or ignorance (pointing, for example, to the fact that white nationalism seems to ignore and marginalize the issue of slavery), attract stock rebuttal responses that had been previously featured on the site many times. For example, we often saw repetition of what Bob Whitaker called "the mantra": that critiques of white nationalists as "racists" were merely criticisms of white people in disguise[1]. We also saw repetition of the idea that many white people were also slaves, and that mainstream arguments about slavery overlooked this. What is critical here, in our view, is not the arguments themselves but the fact that they were repeated over and over again in response to different dissonant challenges by the different members of the site. Indeed, forum moderators at times made explicit reference to the fact that they had allowed opposing views onto the site purely so that common arguments against them could be rehearsed. Arguably one of the most common "echoes" in this echo chamber is the sound of opposing viewpoints being undermined and marginalized.

Second, we identified *selective evidencing* as a further type of counter-argument. This involves selectively incorporating decontextualized news reports, government statistics, and academic articles in order to support some of the fundamental "scientific" claims of white nationalism, especially around the idea that there are genuine differences between races. Posts about differential rates of sickness between races were a strong theme, for example those highlighting a news report suggesting that black people were more likely to have AIDS[2] or contract skin cancer[3]. Sometimes this practice of selective evidencing involved reporting on social problems and suggesting (implicitly or explicitly) that they are related to these genetic differences: for example, statistics showing apparent disproportionate receipt of social security checks by racial minorities[4], or a study showing that multicultural neighborhoods have lower trust[5]. What we found striking in this process was that much of the evidence used was drawn from what might be regarded as 'mainstream' sources (such as major news outlets, government statistics portals and academic institutions), rather than from other white nationalist sites around the internet. Simply by omitting crucial pieces of context, or slightly misreading the intent of the article, members of Stormfront were able to reformat mainstream information in a way that made it consistent with their belief system. This raises serious questions about the value of measuring source exposure as an indicator of homophilous or ideologically extreme communities.

---

[1] See: https://www.adl.org/education/references/hate-symbols/anti-racist-is-a-code-for-anti-white
[2] See http://edition.cnn.com/2008/HEALTH/conditions/07/29/black.aids.report/index.html

**Discussion: Mass Experiences of Dissonance and their Consequences**

This article has investigated the impact of opposing viewpoints on echo chambers. Our main finding is that these viewpoints are not toxic for echo chambers: they don't cause people to stop participation, and they aren't even systematically avoided. Rather, they stimulate lots of discussion, and stimulate people to come back. They are frequently engaged with by the most senior members of the forum, more so than other posts.

It is worth reviewing the consequences of these claims. We identify three of these. First, they suggest that the measurement of cross cutting exposure or source diversity, which is the staple means by which echo chambers are operationalized in the literature, may be misleading. In Stormfront, individuals frequently come into contact with oppositional material (and indeed, as highlighted above, content from mainstream news sites). However, we still think they are, in an important respect, embedded in an echo chamber, because an extreme viewpoint is in the mainstream on the site. Hence work which uses cross cutting exposure as evidence against the existence echo chambers may be misleading. Proposals that individuals should try and 'break out' of echo chambers by listening to alternative viewpoints, however well meaning, may also not have the intended effect (Funnell, n.d.; Grimes, 2017; Nguyen, 2019).

---

[3] See http://news.bbc.co.uk/1/hi/health/5219752.stm
[4] See http://www.census.gov/apsd/www/statbrief/sb95_22.pdf
[5] See: https://www.ft.com/content/c4ac4a74-570f-11db-9110-0000779e2340

A second point is the need to reconsider echo chamber theory as a theory of group behavior. At the moment, by and large, emphasis has been placed on unconscious information processing biases leading to the formation of ideological homogeneity: individuals selectively expose themselves to information they want to see, and selectively avoid other sources. Some of the material we have reviewed here, by contrast, points to a much more deliberate act of echo chamber building by people involved in the forum, which includes continually rebutting dissonant material when it is discovered and providing ample consonant material to be consumed. This makes it worth highlighting the literature on two sided argumentation, which has shown how arguments are more persuasive if they are presented alongside the opposing view (see e.g. O'Keefe, 2002). It also seems worth highlighting again Festinger's early claim that a *mass* experience of dissonance could lead to beliefs becoming stronger, as when members of a group experience dissonance at the same time they enter a mutually reinforcing cycle of dissonance reduction. This seems a very fitting description of what happens on Stormfront (see discussion in Abelson, 1959, p. 346; see also Tormala and Petty, 2004; McKimmie *et al.*, 2003; McKimmie, Terry and Hogg, 2009). The constant exposure to relatively low levels of dissonance may contribute to beliefs becoming progressively stronger.

A final point, related to both of the above, is the means by which echo chambers have power to motivate people to political action. Up until now the literature has emphasized that they give people the ability to ignore or marginalize opposing viewpoints, which then motivates them to take action on behalf of their own group. Our study presents a different potential explanation: the power of echo chambers may derive from the cognitive tools which they provide people to allow them to hold extreme viewpoints *whilst being exposed* to dissonant situations. Indeed, in a way it might be considered surprising the extent to which ideological homogeneity of input has been emphasized in the echo chamber literature. If this was the only way of maintaining extreme beliefs, then such beliefs would arguably be of little practical consequence, as those holding them would, for example, be unable to take part in any political action that involves exposure to opposition. As Anderson and McGuire put it: unchallenged beliefs are known to be "highly vulnerable to persuasive attacks" (1965). By contrast the echo chamber we study here actively stimulates exposure to opposing beliefs. The *echoes* in this echo chamber are the repetition of the cognitive tools and techniques which allow these opposing beliefs to be ignored and marginalized. Hence, the echo chamber can exist in spite of the existence of opposing views; indeed, this may be what causes it to thrive.

## APPENDIX A1: Description of data collection and coding procedures.

*A1.1 Sub-forum description*

**Table A1: Stormfront sub-forums included in the study**

| Sub-forum name | Description (as shown on Stormfront) | Number of Posts Observed (2012 only) | Number of Comments Observed (2012 only) |
|---|---|---|---|
| Opposing Views Forum | "For all our opponents who want to argue with White Nationalists" | 1,468 | 34,368 |
| Events | "White Nationalist demonstrations, rallies, conferences, talk shows, media interviews." | 116 | 844 |
| Local and Regional | "Contact information for those who want to work together in their communities." | 175 | 1344 |
| For Stormfront Ladies Only | "Sugar and spice, and everything nice." | 208 | 5640 |
| Politics & Continuing Crises | "Practical politics, including Ron Paul, Bugsters, Tea Parties, Occupy Wall Street, and, yes, the occasional conspiracy theory." | 1,922 | 23,995 |
| Strategy and Tactics | "Promoting White Rights through local organization." | 139 | 1,077 |
| Talk | "Meet other White Nationalists for romance or friendship." | 236 | 2,087 |

*Note: The descriptions are taken directly from the stormfront.org website (see stormfront.org/forum).*

A description of the sub-forums included in the study can be found in Table A1. The opposing views sub-forum is the main area of interest, where individuals using the forum can encounter both dissonant and consonant material. The other six sub-forums were selected to provide a range of different types of conversation: some oriented towards the ideological and campaigning aspects of white nationalism, others more informal. These six were also selected as together they contained approximately the same amount of comments as the opposing views forum itself during the study period. The values reported in the table refer to 2012 only, which was the main year under study.

*A1.2 Data completeness verification*

Our data was scraped automatically from the Stormfront website, and covers the period 2011 to 2013 (though we only apply our coding to the year 2012). The automatic scraper sought to recursively download threads from the Stormfront website by iterating through pages within these threads and extracting the comments and posts within them into a dataset.

While we believe the scraper performed reasonably well, we also know that intermittent server outages and malformatted HTML on the site may have caused some data quality issues. Therefore, we sought to verify the completeness of the data by comparing it to the actual number of contributions made to our forums of interest during this time period.

We sought these numbers by making use of the Wayback Machine (see https://archive.org/web/). This website archives snapshots of other websites from around the web. It allowed us to look at the state of the Stormfront forum at the start and end of our data capture window (we used Wayback Machine snapshots from the 30th of December 2010 and the 2nd of January 2014, which were the closest available dates to the start and end of our data capture period). In particular, we looked at records of the number of contributions made in each sub-forum of the site. The way these are counted on Stormfront is not completely clear, however in total for our forums of interest we observed 281,341 contributions whilst on the Wayback machine there were 290,691. So overall our total is 96.8% of that reported on Stormfront. A nuance to note is that we actually have slightly more contributions in some of our forums than are listed on the site. It wasn't quite clear why this is and may mean that the Stormfront forum is reporting slightly different statistics. If we disregard these 'overfull' forums from our validation analysis our completeness figure becomes a more conservative 93.2%. This is the figure we have reported in the main text.

### *A1.3 Defining white nationalism*

A definition of white nationalism was critical for our coding procedure to take place. Our definition is based on a close reading of material contributed to the Stormfront website, though works by Meddaugh & Kay (2009) and Bowman-Grieve (2009) also offered considerable inspiration. We define the ideology as consisting of the following basic beliefs.

First, white nationalists believe that there are important differences between 'races' of humans, and that these differences emerge from genetic features. This does not necessarily equate with believing that different races are inferior to the 'white race', though many white nationalists also appear to believe other races are less intelligent and/or more prone to criminality than they are.

Second, white nationalists believe that different nations of the world should be ethnically homogeneous, and as such that white people should be provided with an 'all white' or at least 'majority white' nation in which to live. Opinions diverge considerably about justifiable means for achieving such an outcome, with some advocating peaceful solutions whilst others suggest force would be legitimate.

Finally, white nationalists tend to believe that the white race is in decline or being replaced by policies of multiculturalism and immigration. Some link this to ideas about a (Zionist) conspiracy of global elites.

Needless to say, we think that the beliefs outlined here show the 'extremism' of white nationalist views, and hence their separation from mainstream society.

*A1.4 Coding description*

We chose to code every post (these are original contributions which start discussion threads) in the opposing views forum in the year 2012. The coding was performed by two of the authors. Each author was assigned half of the 1,468 posts to code. The authors read the complete text of the post, and assigned it to one of three categories: 'consonant', 'dissonant' and 'miscellaneous'. The median post was around 540 characters in length, or approximately 100 words.

'Consonant' posts are ones that were broadly ideologically consonant with the belief system of white nationalism as described in appendix A1.3. Postings which expressed support for any of the beliefs we identify as relating to white nationalism were labelled as consonant. Such posts could still be about questions about the best way to live as a white nationalist, or questions about the detail of the belief system. However, they would not be ones which raised major questions about the faith.

'Dissonant' posts, by contrast, are ones which sought to attack or undermine white nationalism in some way. There were many different approaches taken by the authors of dissonant posts. Some sought to highlight what they perceived of as logical fallacies in the white nationalist argument. Others contradicted some of the scientific claims on which the idea of racial differences was based. Also common were postings which involved invective or ad hominem attacks on white nationalists themselves. The main thing these posts have in common is that a white nationalist would feel that their views were, in some way, being challenged.

'Miscellaneous' posts were ones which did not clearly fall into either of the above two categories. These might be posts which were off topic (i.e. did not engage in ideological debate), or ones in which the intent of the poster was not clear.

One hundred of these posts were also selected at random and coded by both of the authors, allowing us to produce an estimate of inter-coder reliability for the coding exercise. We observed an 88% agreement between the authors, which equated to a Krippendorff's Alpha of 0.814, which is usually considered a good level of inter-coder reliability.

**Table A2: Confusion Matrix**

|           | Consonant | Dissonant | Misc |
|-----------|-----------|-----------|------|
| Consonant | 40        | 2         | 3    |
| Dissonant | 4         | 28        | 2    |
| Misc      | 1         | 0         | 20   |

The confusion matrix for the coding is shown in Table A2 above. There was disagreement on 12 of the items coded, of which 6 were disagreements between the key categories of interest (i.e. posts where one coder marked 'Consonant' and one marked 'Dissonant'). The balance of sample sizes between all the three categories is reasonable. Overall, we feel the results support the validity of the coding scheme.

## *A1.5 Descriptive Statistics*

## Table A3: Descriptive Statistics

|  | **Users** |
|---|---|
| Members | 3,324 (77%) |
| Non-members | 1,008 (23%) |
| *Total users* | 4,332 |

|  | **Threads** | **Comments** |
|---|---|---|
| Inside Opposing Views | 1,468 | 34,368 |
| Outside Opposing Views | 2,796 | 34,987 |
| *Total* | 4,264 | 69,335 |

| Consonant Threads | 702 |
|---|---|
| Dissonant Threads | 469 |
| Miscellaneous | 297 |

|  | Median | Mean |
|---|---|---|
| **Seniority** | 155 | 200 |
| **Post length (characters)** | 538 | 1,111 |

## APPENDIX A2: Model diagnostics

### *A2. 1 Diagnostics for post response and contributor seniority models*

Models 1.0 – 2.2 are multilevel models. Data was grouped by the user creating the post, as we might expect different users to generate different typical reactions when posting threads, and by the month of posting, as we might expect response levels to fluctuate over time. We did not use the forum of posting as a grouping factor because variation here is already largely accounted for by the post type categorical variable.

The following diagnostic checks were performed. The normality of the dependent variable was inspected (leading it to be log transformed). Plots of residuals versus fitted values were inspected. These suggested no problems in the case of Models 1.0 – 1.2, though some skew was observed in the case of Models 2.0 – 2.2. This skew was caused by posts with few responses. The dependent variable of these models is the average seniority of people commenting on a post: clearly, posts which had fewer comments would be more likely to have higher variance in the average seniority of responders. Further models were produced with posts with less than ten responses removed. The fit was much improved, but the results were largely the same.

The normality of residuals was also inspected. This suggested some problems in the case of models 1.0 – 1.2 which were resolved by removing posts with zero responses (the results of these new models being substantially the same as the ones reported in the main text). These plots also suggested the same truncation of posts with few responses for Models 2.0 – 2.2 which was already performed above. Finally, hat values were calculated to check for high leverage points. However, no hat values greater than the conventional cut-off of 0.4 were observed.

Overall the diagnostics provided no reason to doubt the results of the models reported in the main text.

*A2.2 Gap Time Analysis Diagnostics*

Models 3.0 and 3.1 are a type of proportional hazards model known as a recurrent events model (specifically a PWP Gap Time model). These models function as a normal event history analysis, except that there are multiple events per individual. The model hence clusters standard errors by individual, and also stratifies observations by the number of previous events and individual has had. There are three main diagnostic considerations for such a model. First, the number of individuals can be relatively low at higher strata, which can lead to instability in estimates (see Amorim & Cai, 2014). Hence, truncated models were produced that disregarded events after the 169$^{th}$ event (which was the largest strata with at least 50 individuals within it). The results of these models were substantially the same. Second, the models were checked for influential observations, measured using DF Beta scores. No observations above the conventional cutoff were observed (the cutoff being $2 / \sqrt{n}$).

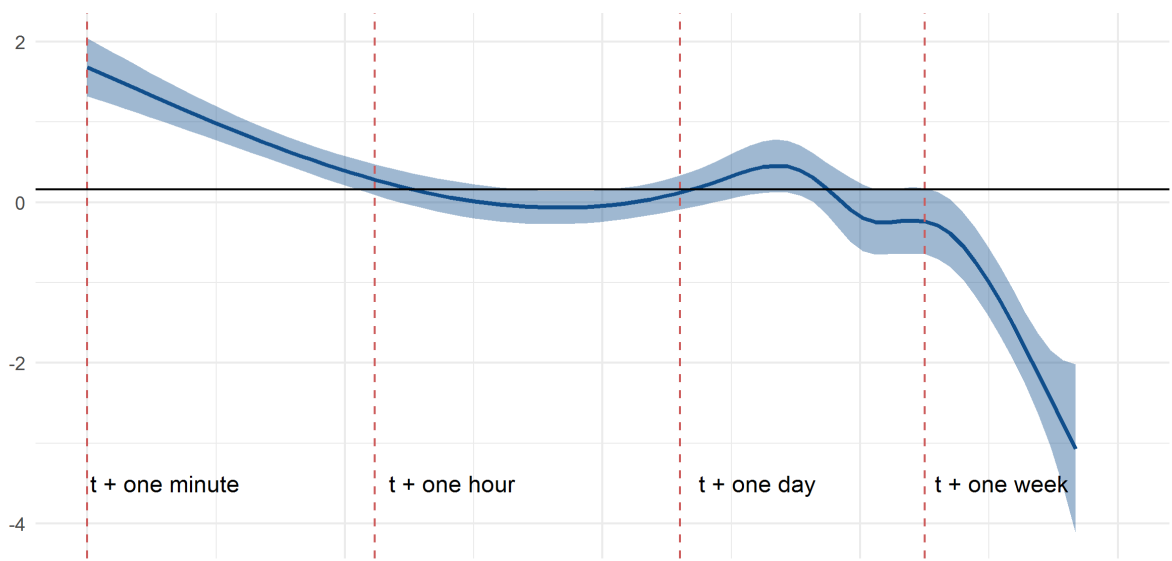

**Figure A1: Trend line indicates a smoothed LOESS estimation of the evolution of the main effect from Model 3.0 over time.** *Note: The black horizontal line indicates the estimate reported in Model 3.0.*

Finally, the proportional hazards assumption was checked for both models. This was found to be violated, meaning that the ratio of hazards between groups is not constant over time. The evolution of the estimate, compared to the main effect reported in Model 3.0 (the solid black line) is shown in figure A1 above. The figure shows that the effect is strongly positive for around the first hour after posting. After this time, it fluctuates around zero until around a week, when it turns negative. This indicates that the main 'stimulus' effect of encountering an opposing view occurs shortly after having engaged with it. This is reported in the main text.

### *A2.3 Markov chain full transition probabilities*

**Table A3: Markov Chain Transition Probabilities**

| Previous Contribution | Next Contribution | Estimate | Lower Bound | Upper Bound |
| --- | --- | --- | --- | --- |
| consonant | consonant | 0.5011 | 0.487 | 0.5153 |
| consonant | outside | 0.2045 | 0.1954 | 0.2135 |
| consonant | strong | 0.2171 | 0.2078 | 0.2264 |
| consonant | unknown | 0.0773 | 0.0718 | 0.0829 |
| outside | consonant | 0.0539 | 0.0515 | 0.0563 |
| outside | outside | 0.8555 | 0.846 | 0.865 |
| outside | strong | 0.0665 | 0.0638 | 0.0691 |
| outside | unknown | 0.0241 | 0.0225 | 0.0257 |
| strong | consonant | 0.1991 | 0.1907 | 0.2076 |
| strong | outside | 0.2299 | 0.2208 | 0.2389 |
| strong | strong | 0.4948 | 0.4815 | 0.5081 |
| strong | unknown | 0.0762 | 0.071 | 0.0814 |
| unknown | consonant | 0.1661 | 0.1537 | 0.1784 |
| unknown | outside | 0.1984 | 0.1849 | 0.2119 |
| unknown | strong | 0.2194 | 0.2052 | 0.2336 |
| unknown | unknown | 0.4161 | 0.3966 | 0.4357 |

Table A3 above provides the full transition probabilities for all possible pairs of states in our Markov chain model. The probabilities reflect our estimate of the next type of discussion thread someone is likely to contribute to, given the previous one they contributed to. For example, if an individual contributed to a consonant discussion thread within the opposing views forum, we estimate a 0.5011 probability that their next contribution will also be a consonant discussion thread in the opposing views forum.